\documentclass[aps,prb,twocolumn,showpacs,floatfix,superscriptaddress]{revtex4-1}
\usepackage{amssymb}
\usepackage{amsmath}
\usepackage{graphicx}
\usepackage{dcolumn}
\usepackage{bm}
\usepackage{epsfig}
\usepackage{color}

\begin{document}

\title{Excited states of exciton-polariton condensates in 2D and 1D harmonic traps}

\author{C. Trallero-Giner}
\affiliation{Facultad de F\'{\i}sica, Universidad de La Habana, Vedado 10400, La Habana,
Cuba}
\author{M. V. Durnev}
\affiliation{Spin Optics Laboratory, State University of St-Petersburg, 1, Ulianovskaya,
St-Petersburg, 19850, Russia }
\affiliation{Ioffe Physical-Technical Institute of the RAS, 194021 St.-Petersburg, Russia}
\author{Y. N\'{u}\~{n}ez Fern\'{a}ndez}
\affiliation{Centro At\'{o}mico de Bariloche, 8400, Argentina}
\author{M. I. Vasilevskiy}
\affiliation{Centro de F\'{\i}sica and Departamento de F\'{\i}sica, Universidade do
Minho, Campus de Gualtar, Braga 4710-057, Portugal}
\author{V. L\'{o}pez-Richard}
\affiliation{Departamento de Fisica, Universidade Federal de S\~{a}o Carlos, 13.565-905, S%
\~{a}o Carlos, S\~{a}o Paulo, Brazil}
\author{A. Kavokin}
\affiliation{Physics and Astronomy School, University of Southampton, Highfield,
Southampton, SO17 1BJ, UK and Spin Optics Laboratory, State University of
St-Petersburg, 1, Ulianovskaya, St-Petersburg, 198504, Russia}

\begin{abstract}
We present a theoretical description of Bogolyubov-type excitations of
exciton-polariton Bose-Einstein condensates (BECs) in semiconductor
microcavities. For a typical two dimensional (2D) BEC we focus on two
limiting cases, the weak- and strong-coupling regimes, where a perturbation
theory and the Thomas--Fermi approximation, respectively, are valid. We
calculate integrated scattering intensity spectra for probing the collective
excitations of the condensate in both considered limits. Moreover, in
relation to recent experiments on optical modulation allowing localization
of condensates in a trap with well controlled shape and dimensions, we study
the quasi-one dimensional (1D) motion of the BEC in microwires and report
the corresponding Bogolyubov's excitation spectrum. We show that in 1D case
the characteristic polariton-polariton interaction constant is expressed as $%
g_{1}=3\lambda \mathcal{N}/(2L_{y})$ ($\lambda $ is the 2D
polariton-polaritons interaction parameter in the cavity, $\mathcal{N}$ the
number of the particles, and $L_{y}$ the wirecavity width). We reveal some
interesting features for 2D and 1D Bogolyubov spectra for both repulsive $%
(\lambda >0)$ and attractive $(\lambda <0)$ interaction.
\end{abstract}

\maketitle

\section{Introduction}

The rich picture of the exciton-polariton Bose-Einstein condensates (BECs)
in semiconductor microcavities has opened the opportunity for exploring a
great variety of phenomena, such as superfluidity of quantum fluid~\cite%
{amo,Keeling}, vortices~\cite{Sarchi}, persistent currents~\cite{Sanvitto},
half-quantum vortices~\cite{Lagoudakis}, as well as applications in quantum
cascade laser~\cite{Liew} and interferometric devices (see Ref.~%
\onlinecite{tosi} and references therein). The scheme of controlling the
dynamic of condensates is ruled by their elementary excitations. The
knowledge of the properties of these (Bogolyubov-type) excitations would
reveal the main causes and allow for the detailed understanding of the
physical phenomenon under consideration. Phonon-like excitation spectrum in
the low-momentum regime of a quantum fluid condensate was firstly
theoretically studied by Bogolyubov.~\cite{Bogoliubov} Finally, M. H.
Anderson and co-workers~\cite{Anderson} observed the Bogolyubov-de Genes
spectrum in a ultra-cold dilute atom cloud.

Polariton-polariton interaction and the behavior of the excitation spectra
play a fundamental role for understanding the underlying physics of the BEC
dynamic in semiconductor microcavities. Utsunomiya \textit{et.al} have
realized~\cite{Utsunomiya} the first experimental observation of Bogolyubov
spectrum in a GaAs/AlGaAs microcavity by showing, in the phonon-like regime,
a clear linearization of the quadratic polariton dispersion as a function of
the in-plane wavevector \textbf{k. }Nevertheless\textbf{, }the
exciton-polariton excitation energy is modified if the condensate is loaded
in a 2D trap potential, thus the spectrum must be characterized by
appropriate quantum numbers describing the confined phonon-like excitations.

In addition to the two dimensionality nature of the BEC in semiconductor
microcavities, a dynamical condensation of exciton-polaritons can also be
induced in one-dimensional (1D) systems. Nowadays, it is possible to realize and
manipulate diverse trap potentials for exciton-polaritons.~\cite%
{tosi,Bajoni,Wertz,Amo1} This experimental ability enhances the range of
potential applications, such as the design of condensate circuits,~\cite%
{tosi} among others. In particular, 1D parabolic confined potentials can be
generated in microwires or by optical manipulation. Then, optical modulation
allows to control the shape of the condensate wave functions and to study
their quasi-1D motion in the microcavity. \cite{Wertz,tim} In particular,
exciton-polariton 1D harmonic traps have been induced by employing two pump
laser beams. \cite{tosi,KavovinPolariton}

The aim of this paper is to give a mathematical description of elementary
excitations of a confined BEC\ in microcavities, with a special emphasis on
the shape and dimensionality of the trap potentials. We explore convenient
analytical descriptions of the condensate that can be used to study its
dynamics and related physical phenomena, such as vortices, persistent
currents and superfluidity.

This manuscript is organized as follows. Section II is devoted to the
Bogolyubov-type elementary excitations in a 2D parabolic potential. Two
approaches are discussed: (i) a perturbative method where the cubic term
present in the non-linear Gross-Pitaevskii equation (GPE), can be considered
as perturbation in comparison to the trap potential, and (ii) the so-called
strong coupling regime or Thomas-Fermi limit where the collective
excitations of the ground state are described by the variation of its
density. Also, we report calculated spectra of electromagnetic waves
scattered by the condensate comparing both limits of weak and strong
non-linear interactions. In Sec. III, we first deal with the reduction of
the spatial 2D GPE to an effective 1D equation by "freezing" the transversal
direction if the cavity width is much smaller than the harmonic oscillator
length. The procedure allows us to rigorously obtain the effective 1D
polariton-polariton interaction constant. Then, in the framework of the
present model, we derive the Bogolyubov excitations for the 1D parabolic and
semi-parabolic traps and show the main differences between them. Finally, in
Sec. IV is devoted to conclusions.

\section{Two dimensional collective excitations}

Within the framework of the mean field theory, the physical characteristics
of a trapped BEC described by a macroscopic wave function $\Psi (\mathbf{r}%
,t)$ are ruled by the time dependent 2D nonlinear GPE,

\begin{equation}
i\hbar \partial _{t}\Psi =\left( -\frac{\hbar ^{2}}{2m_{\ast }}\Delta +V(%
\mathbf{r})+\lambda \left\vert \Psi \right\vert ^{2}\right) \Psi \text{ },
\label{eq:GPT}
\end{equation}%
where $\bm r=(r,\theta )$ is the radius vector in polar coordinates, $%
\lambda $ is the self-interaction parameter,~\cite{Ciuti,prb} $m_{\ast }$ is
the exciton-polariton mass, and $V(\mathbf{r})=m_{\ast }(\omega
_{0x}^{2}x^{2}+\omega _{0y}^{2}y^{2})/2$ is the two dimensional harmonic
potential characterized by the trap frequencies $\omega _{0x},\omega _{0y}$.

The collective excitations can be obtained by applying a small deviation
from the stationary solutions $\Psi _{0}(\mathbf{r},t)=\psi _{0}(\mathbf{r}%
)\exp (-i\mu t/\hbar )$ of Eq.~(\ref{eq:GPT}) in the form of~\cite%
{pitaevskii}

\begin{multline}
\Psi (\mathbf{r},t)=\exp (-i\mu t/\hbar )\left[ \psi _{0}(\mathbf{r})+u(%
\mathbf{r})\exp (-i\omega t)\right. \\
\left. +v^{\ast }(\mathbf{r})\exp (i\omega t)\right] \text{ },
\label{eq:wafeper}
\end{multline}%
where $\mu $ is the chemical potential, $u$ and $v$ are the amplitudes of
the excitation mode with frequency $\omega $. The perturbative nature of the
last two terms in Eq.~(\ref{eq:wafeper}) is ensured if the following conditions are satisfied, $%
\left\langle u|u\right\rangle ,\left\langle v|v\right\rangle \ll
\left\langle \psi _{0}|\psi _{0}\right\rangle $. After substituting the
perturbed wave function (\ref{eq:wafeper}) into Eq.~(\ref{eq:GPT}) and
performing the linearization procedure, one obtains the following eigenvalue
problem for the frequencies $\omega $ and amplitudes $u$ and $v$:
\begin{equation}
\left[
\begin{array}{cc}
-\frac{\hbar ^{2}}{2m_{\ast }}\Delta +U(\mathbf{r}) & \lambda \left\vert
\psi _{0}\right\vert ^{2} \\
&  \\
-\lambda \left\vert \psi _{0}\right\vert ^{2} & \frac{\hbar ^{2}}{2m_{\ast }}%
\Delta -U(\mathbf{r})%
\end{array}%
\right] \left(
\begin{array}{c}
u \\
\\
v%
\end{array}%
\right) =\hbar \omega \left(
\begin{array}{c}
u \\
\\
v%
\end{array}%
\right) \text{ },  \label{eq:bogoliubov}
\end{equation}%
with
\begin{equation}
U(\mathbf{r})=V(\mathbf{r})+2\lambda \left\vert \psi _{0}(\mathbf{r}%
)\right\vert ^{2}-\mu \text{ }.  \label{eq:potential}
\end{equation}%
The operator in Eq.~(\ref{eq:bogoliubov}) is not Hermitian, however, its
spectrum lies entirely in real space~\cite{you1997} with the set of positive
and negative values $\omega $ corresponding to $(u:v)^{T}$ and $(v^{\ast
}:u^{\ast })^{T}$ states, respectively. The ground state characteristics $%
\psi _{0}$ and $\mu $ can be found from the stationary GPE
\begin{equation}
\left( -\frac{\hbar ^{2}}{2m_{\ast }}\Delta +V(\mathbf{r})+\lambda
\left\vert \psi _{0}\right\vert ^{2}\right) \psi _{0}=\mu \psi _{0}\text{ },
\label{eq:ground}
\end{equation}%
with the boundary conditions $\psi _{0}\rightarrow 0$ at $r\rightarrow
\infty $ and normalized over the total number of condensed particles, $%
\mathcal{N},$
\begin{equation}
\left\langle \psi _{0}\left\vert {}\right. \psi _{0}\right\rangle =\int
\left\vert \psi _{0}(\mathbf{r})\right\vert ^{2}d\mathbf{r}=\mathcal{N}\text{
}.  \label{eq:normalize}
\end{equation}

Let us now introduce the dimensionless parameter $\Lambda =\lambda m_{\ast }%
\mathcal{N}/\hbar ^{2}$ which describes the strength of interaction in the
system. We will further focus on the two important limiting cases when
solutions of Eqs.~( \ref{eq:bogoliubov}) and (\ref{eq:ground}) can be found
analytically, namely, the limits of sufficiently small and large values of $%
\Lambda $. In the former limit the interaction term can be treated in the
framework of the perturbation theory while in the latter case (the so-called
Thomas-Fermi approximation) collective excitations are the solutions of the
hydrodynamic-like equations.~\cite{stringari1996}

\subsection{Perturbative method}

As known, the polariton-polariton self-interaction parameter $\lambda $
depends on exciton-photon detuning $\delta $.~\cite{Ciuti} For typical
GaAs/AlGaAs microcavities, the perturbation theory for the GPE (\ref%
{eq:ground}) is valid if the number of particles in the condensate $\mathcal{%
N}\leq 10^{4}$ for -10 meV$<\delta <$3 meV or $\mathcal{N}\leq 10^{6}$ if
the detuning lies in the interval 3 meV$<\delta <$7 meV.~\cite{prb2} All
these cases correspond to the dimensionless self-interaction coefficient's
values -3$<\Lambda <$3.~\cite{prb} In this range of parameter $\lambda ,$
the nonlinear term $\lambda \left\vert \psi _{0}\right\vert ^{2}$ in Eq.~(%
\ref{eq:ground}) can be considered as a small perturbation with respect to
the isotropic ($\omega _{0x}=\omega _{0y}=\omega _{0}$) harmonic trap
confinement potential, $m_{\ast }\omega _{0}^{2}r^{2}/2.$\ Hence, the order
parameter $\psi _{0}$ can be expanded in series of the complete set of 2D
harmonic oscillator wave functions~\cite{prb2}

\begin{equation}
\varphi _{N,m_{z}}(\rho ,\theta )=\frac{\exp (im_{z}\theta )}{\sqrt{2\pi }}%
R_{N,m}(\rho )\text{ },
\end{equation}%
with $\rho =r/a,$ $a=\sqrt{\hbar /m_{\ast }\omega _{0}}$ the characteristic
unit length, $m_{z}$ the $z$-projection of the angular momentum, $%
m=\left\vert m_{z}\right\vert $, and $N=0,1,2,...$ $.$ The corresponding
energies measured in units of $\hslash \omega _{0}$ are $\epsilon _{N}=N+1$.

The chemical potential and the condensate distribution $n_{0}=\left\vert
\psi _{0}\right\vert ^{2}$ up to the second and first order in $\Lambda $
read~\cite{prb}
\begin{equation}
\frac{\mu }{\hbar \omega _{0}}=1+\frac{\Lambda }{2\pi }-\frac{3\Lambda ^{2}}{%
8\pi ^{2}}\ln (4/3)\text{ },  \label{eq:groundstate_mu1}
\end{equation}

\begin{eqnarray}
n_{0}(r) &=&\frac{\overline{n_{0}(\rho )}}{a^{2}}=\frac{n^{(0)}+\Lambda
n^{(1)}}{a^{2}}  \label{eq:groundstate_psi1} \\
&=&\frac{1}{\pi a^{2}}\exp \left( -\rho ^{2}\right) \left[ 1+\Lambda F(\rho )%
\right] \text{ },  \notag
\end{eqnarray}%
with

\begin{equation}
F(\rho )=\dfrac{1}{2\pi }\left[ \gamma +\ln \left( \rho ^{2}/2\right)
+\Gamma \left( 0,-\rho ^{2}\right) \right] \text{ }.
\end{equation}%
Here $\Gamma (0,z)$ is the incomplete gamma function and $\gamma $ is the
Euler-Mascheroni constant.~\cite{GRTable} Since the perturbation, $\lambda
\left\vert \psi _{0}\right\vert ^{2},$ in Eq.~(\ref{eq:bogoliubov}) does not
mix states with different angular momenta $m=\left\vert m_{z}\right\vert ,$
we can search the required solutions as $u[v]=\exp (im_{z}\theta
)u_{N,m}(\rho )[v_{N,m}(\rho )]/\sqrt{2\pi }$ with the amplitudes $u_{N,m}$
and $v_{N,m}$ expanded over the set of 2D radial components of oscillator
wave functions $\left\{ R_{N,m}\right\} ,$

\begin{equation}
\left(
\begin{array}{c}
u_{N,m} \\
v_{N,m}%
\end{array}%
\right) =\sum\limits_{N_{1}=0}^{\infty }\left(
\begin{array}{c}
A_{NN_{1}}R_{N_{1},m} \\
B_{NN_{1}}R_{N_{1},m}%
\end{array}%
\right) \text{ }.  \label{serie1}
\end{equation}%
Substituting Eq.~(\ref{serie1}) into Eq.~(\ref{eq:bogoliubov}), one can
obtain the spectrum of the Bogolyubov's excitations (see Appendix A for
details). It is possible to show that $\omega _{N,m}$ up to the second order
in $\Lambda $ can be cast as
\begin{equation}
\omega _{N,m}=\omega _{0}\left( N+\Lambda \varpi _{N,m}^{(1)}+\Lambda
^{2}\varpi _{N,m}^{(2)}\right) \text{ ,}  \label{freque}
\end{equation}%
where the coefficients $\varpi _{N,m}^{(1)}$ and $\varpi _{N,m}^{(2)}$ are
obtained in the Appendix A.

\begin{figure}[tbp]
\includegraphics[width = 0.48\textwidth]{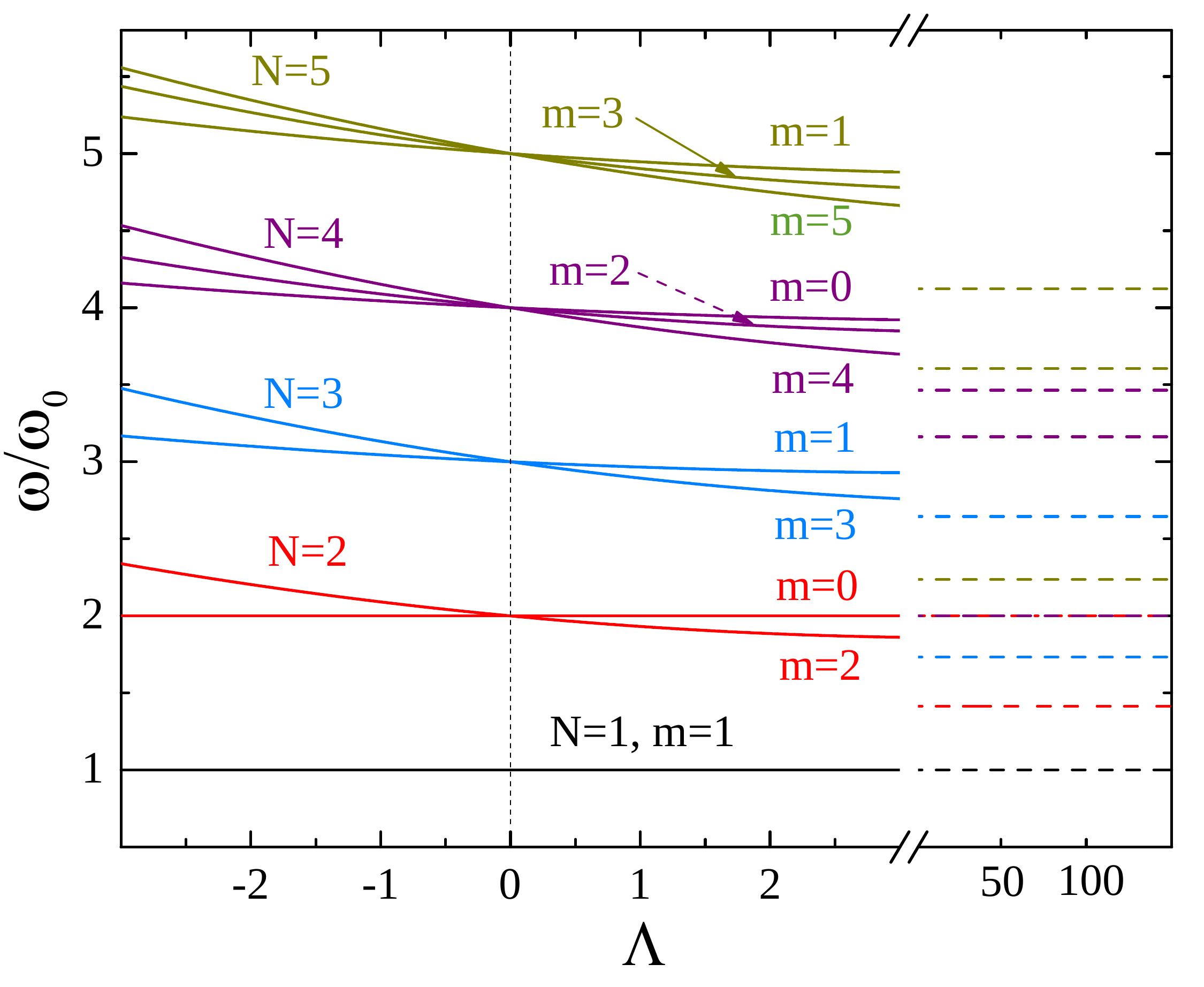}
\caption{(Color online): Excitation spectrum $\protect\omega _{N,m}$ as a
function of the reduced self-interaction parameter $\Lambda $. Solid lines:
Bogolyubov's spectrum solution of Eq.~(\protect\ref{freque}). Dashed lines:
Thomas-Fermi approximation after Eq.~(\protect\ref{TF}).}
\label{fig:fig1}
\end{figure}
\begin{figure}[tbh]
\includegraphics[width = 0.45\textwidth]{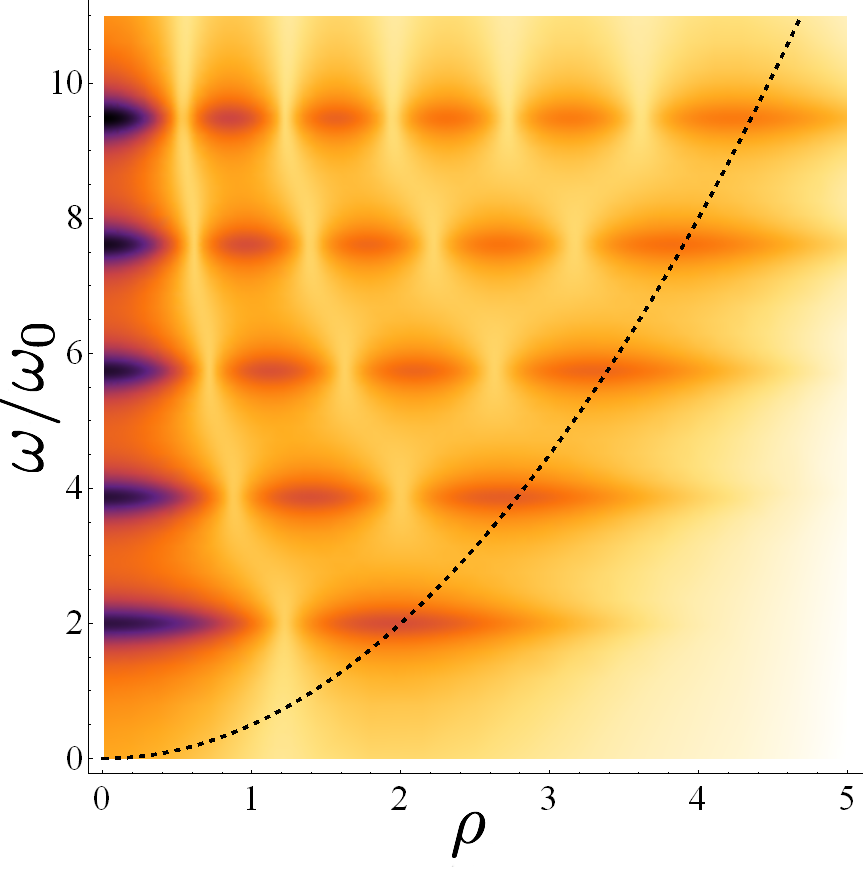}
\caption{(Color online): Calculated frequency-coordinate space distribution
function, $D_{exc}(\mathbf{r};\protect\omega )$, of the Bogolyubov
collective excitation (see text) in a 2D parabolic trap. The parameter used
are: $m=0$, $\Lambda =1$ and $\protect\gamma _{d}=0.1\protect\omega _{0}$.
The dark dot curve shows the dimensionless harmonic potential $V(r)/\protect%
\omega _{0}=\protect\rho ^{2}/2$.}
\label{fig:fig2}
\end{figure}

Figure~\ref{fig:fig1} displays the excitation frequencies $\omega _{N,m}$
for the lowest 11 collective modes \textit{versus} the dimensionless
interaction parameter $\Lambda $. Let us note that the excitation
frequencies in the non-interacting limit are equal to those of the
two-dimensional harmonic oscillator measured from the zero-level $\omega
_{0} $ oscillations. At $\Lambda =0,$ the system is $(N+1)$ degenerate with
respect to angular momentum value $m_{z},$ which is a direct consequence of
the axial symmetry inherent to the isotropic 2D harmonic confinement
potential. At $\Lambda \neq 0,$ the frequency corrections $\varpi
_{N,m}^{(1)}$ and $\varpi _{N,m}^{(2)}$ depend on $m$ (see Eqs.~(\ref%
{lambda1}) and (\ref{lambda2}), respectively). Hence, the non-linear
perturbative term splits the energy spectrum by the absolute value of the
angular momenta projection $m,$ leaving the two-fold degeneracy with respect
to the sign of $m_{z}$. Figure~\ref{fig:fig1} shows that for the repulsive
interaction, the excitation energies decrease with the increase of the
condensate density. From the figure it follows that the doubly degenerate
dipole excitation states $N=1,m_{z}=\pm 1$ (mode $\omega _{1,1})$ are
unaffected by the nonlinear interaction. This mode is harmonic, in
agreement with the Kohn's theorem,~\cite{Kohn} and it represents a rigid
motion of the center of mass.~\cite{stringari1996, edwards1996} Notice that
the same result holds under Thomas-Fermi limit in the framework of the
hydrodynamic approximation as it is represented by dashed lines in Fig.~\ref%
{fig:fig1}. For the attractive potential, the excitation energy $\omega
_{N,m}(\Lambda <0)$ increases as $\Lambda $ decreases and, for a given radial
quantum number $N$, the lower value of $\omega _{N,m}(\Lambda <0)$
corresponds to the lower quantum number $m.$ Notice that in the case of
repulsive interaction the opposite trend is obtained, i.e. $\omega _{N,m}$
decreases as $\Lambda $ increases and $\omega _{N,m}(\Lambda >0)>$ $\omega
_{N,m+2}(\Lambda >0).$ The amplitudes $u_{N,m}$ and $v_{N,m}$ up to first
order in $\Lambda $ are given in Appendix A. For a clearer demonstration of
the space distribution of excitations, we evaluate the density function
excited particles distribution $D_{exc}=-\emph{Im}\left\{ Tr\left[ \hat{G}(%
\mathbf{r},\mathbf{r};\omega )\right] \right\} ,$ where $\hat{G}(\mathbf{r},%
\mathbf{r};\omega )$ is the matrix Green's function of the Eq.~(\ref%
{eq:bogoliubov}) given by
\begin{equation}
\hat{G}(\mathbf{r},\mathbf{r}^{\prime };\omega )=\frac{1}{\hbar }%
\sum\limits_{N,m_{z}}\frac{\left[ \hat{\chi}_{N,m_{z}}(\bm{r})\right] ^{T}%
\hat{\chi}_{N,m_{z}}^{\ast }(\bm{r}^{\prime })}{\omega -\omega _{N,m}+%
\mathrm{i}\gamma _{d}}\text{ }.  \label{eq:Green}
\end{equation}%
Here $\hat{\chi}_{N,m_{z}}(\bm{r})=\left[ u_{N,m}(r)\mathrm{e}^{\mathrm{i}%
m_{z}\theta }:v_{N,m}(r)\mathrm{e}^{\mathrm{i}m_{z}\theta }\right] $ and $%
\gamma _{d}$ is the damping parameter of the photon mode in the microcavity.
The summation in Eq.~(\ref{eq:Green}) is performed only over $N$ while the
value of $m_{z}$ is set constant, $m_{z}=0$. In Fig.~\ref{fig:fig2} the
calculated exciton-polariton density distribution $D_{exc}(\mathbf{r};\omega
)$ is shown for $\Lambda =1$, and $\gamma _{d}=0.1\omega _{0}$. The bright
spots observed in the figure correspond to the minima of the excitation density
and are linked to the zeros of the radial density function $\left\vert
R_{N,m=0}(\rho )\right\vert ^{2}$, i.e. for a given $\omega _{N,m=0}$ the
function $D_{exc}$ has $n=N/2$ minimum in space domain.

\subsection{Current density}

Superfluidity, the formation of quantized vortices,~\cite{abo} and
persistent currents~\cite{Sanvitto}\ are among the most interesting
properties of exciton-polariton condensates. Collective fluid dynamics of
condensates can be driven coherently by the continuous-wave pump energy and
triggered by a short pulse of another laser.~\cite{amo} Also, vortices can
be excited by a pulsed probe transferring angular momentum resonantly with
the pumping signal.~ \cite{Sanvitto} In all considered effects, and in a
general sense, assessing the current density, $\mathbf{j}(\mathbf{r},t)%
\mathbf{,}$ becomes necessary and the evolution of the density profile for the
collective excitations is of interest. Starting from the well known equation
for the current density,
\begin{equation}
\mathbf{j}_{N,m_{z}}=\frac{\hslash }{2mi}\left[ \Psi _{N,m_{z}}^{\ast
}\nabla \Psi _{N,m_{z}}-\left( \nabla \Psi _{N,m_{z}}^{\ast }\right) \Psi
_{N,m_{z}}\right] \text{ }  \label{j}
\end{equation}%
and rewriting the wave function as $\Psi _{N,m_{z}}(\mathbf{r},t)=\left\vert
\Psi _{N,m_{z}}(\mathbf{r},t)\right\vert \exp \left( iS_{N,m_{z}}(\mathbf{r}%
,t)\right) $ we obtain:
\begin{equation}
\mathbf{j}_{N,m_{z}}(\mathbf{r},t)=c_{N,m_{z}}(r,t)\mathbf{v}_{s}(\mathbf{r}%
,t)\text{ },  \label{jA}
\end{equation}%
where $\mathbf{v}_{s}(\mathbf{r},t)=\frac{\hslash }{m}\nabla S(\mathbf{r},t)$
is known as the superfluid velocity. Notice that integer vortices are
described by rotation of $S\rightarrow S+2\pi p,$ with $p=\pm 1,\pm 2,....$
and half vortices correspond to $p=\pm 1/2,\pm 3/2,...$~\cite{Lagoudakis} In Eq.~(\ref%
{jA}), $c_{N,m_{z}}(r,t)=\left\vert \Psi _{N,m_{z}}\right\vert
^{2}=\left\vert \psi _{0}(\mathbf{r},t)+\delta \Psi _{N,m_{z}}(\mathbf{r}%
,t)\right\vert ^{2}$ represents the total concentration of particles in the
excited state ($N,m_{z})$.

The non-zero $\theta $-component of the current density associated with an
elementary excitation is obtained by evaluating the gradient,
\begin{equation}
j_{N,m_{z}}^{\theta }(\mathbf{r},t)=\frac{m_{z}\hbar }{2m_{\ast }}\delta
c_{N,m_{z}}(r,t)\text{ },  \label{jtheta}
\end{equation}%
where the condensate density perturbation, $\delta c_{N,m_{z}}$, is given by:

\begin{multline}
\delta c_{N,m_{z}}=\left\vert \Psi _{N,m_{z}}(\mathbf{\rho },t)\right\vert
^{2}-\left\vert \psi _{0}(\mathbf{\rho },t)\right\vert ^{2}=  \label{dc} \\
\sqrt{\frac{2}{\pi }}\cos (\omega _{N,m}t-m_{z}\theta )\left[ \psi
_{0}R_{N,m}+\frac{2}{\sqrt{\pi }}\Lambda \exp \left( -\rho ^{2}/2\right)
\times \right. \\
\sum_{N_{2}\neq N}C_{N,N_{2},m}^{(0)}R_{N_{2},m}\frac{N+3N_{2}}{%
(N-N_{2})(N+N_{2})}\text{ }.
\end{multline}

\begin{figure}[tbp]
\includegraphics[width = 0.45\textwidth]{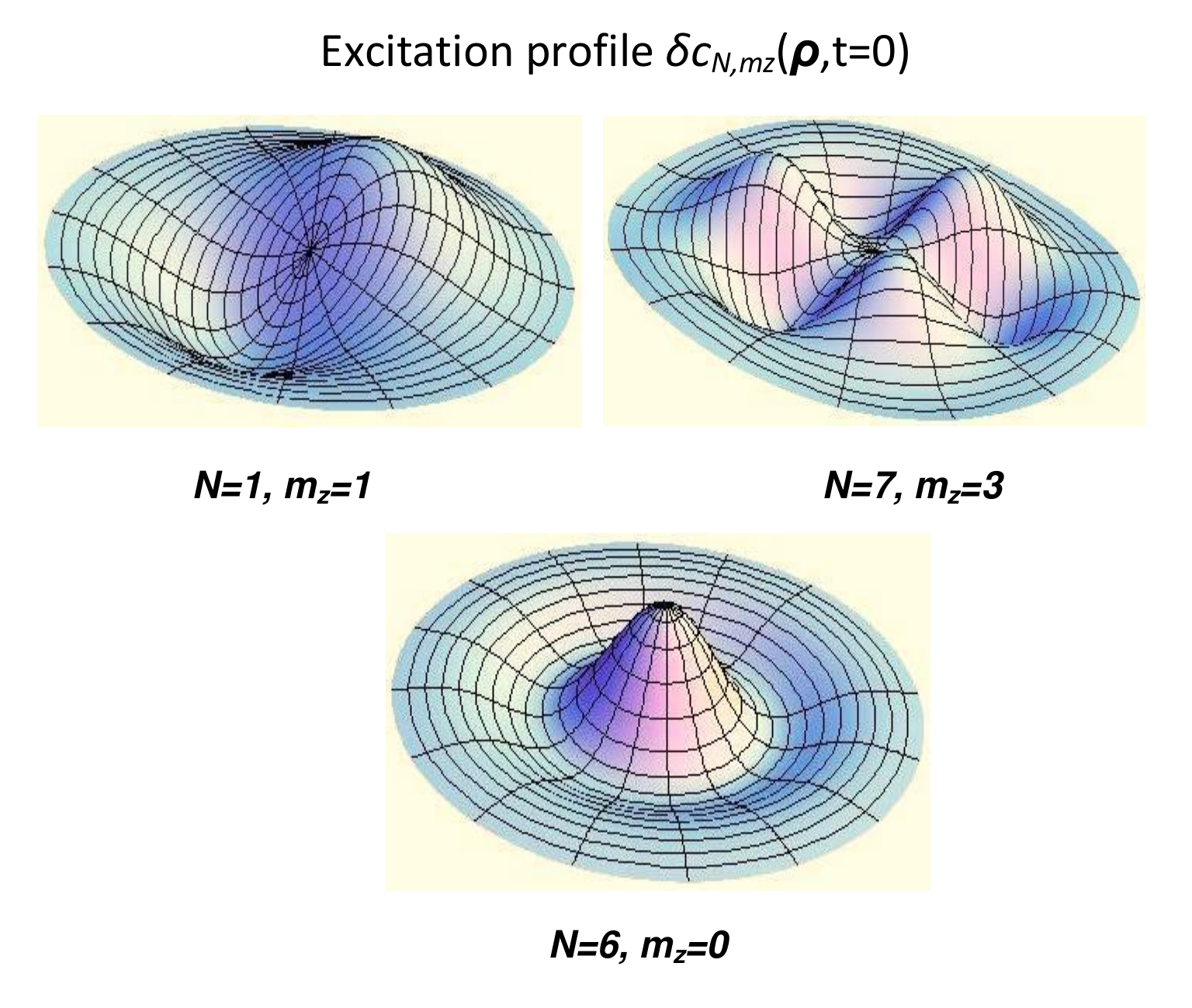}
\caption{(Color online): Excitation profile $\protect\delta c_{N,m}$ given
by Eq.~(\protect\ref{dc}) at $t=0$ for the excited states $(N=1,m_{z}=1)$, $%
(N=7,m_{z}=3)$ and $(N=6,m_{z}=0)$. Here $\Lambda =1$.}
\label{fig:fig3}
\end{figure}

The excitation density profiles $\delta c_{N,m}$ at $t=0$ are displayed in
Fig.~\ref{fig:fig3} for ($N=1,m_{z}=1)$, ($N=7,m_{z}=3),$ and ($%
N=6,m_{z}=0). $ Figure~\ref{fig:fig3} illustrates the symmetry properties of
the excited states as a function of the quantum number $m_{z}.$ Since $%
\delta c_{N,m}(t=0)\sim \cos (m_{z}\theta )$ the profile shows the nodal
distribution at $\theta _{p}=\left( 2p+1\right) \pi /(2m_{z})$ with $%
p=0,1,..,\left\vert m_{z}\right\vert .$

\subsection{Thomas-Fermi limit}

For sufficiently large values of parameter $\Lambda ,$ the density profile
of the condensate becomes smooth enough to omit the kinetic energy term in
Eq.~(\ref{eq:ground}).~\cite{nota0} We then arrive to the so-called
Thomas-Fermi limit with the ground state density,
\begin{equation}
n_{0TF}(\mathbf{r})=%
\begin{cases}
\dfrac{\mu _{TF}}{\hbar \omega _{0}}\dfrac{1}{\Lambda }\left( 1-\dfrac{r^{2}%
}{r_{0}^{2}}\right) , & r<r_{0}\text{ }; \\
0, & r\geq r_{0}\text{ }.%
\end{cases}
\label{eq:ground_tf}
\end{equation}%
Here $r_{0}=\sqrt{2\mu _{TF}/(m_{\ast }\omega _{0}^{2})}$ is the radius of
the condensate ground state in the Thomas-Fermi limit. The normalization
condition~(\ref{eq:normalize}) yields for the chemical potential $\mu
_{TF}=\hbar \omega _{0}\sqrt{\Lambda /\pi }$ so that $\mu _{TF}$ is large
compared to the oscillator energy. For the collective excitations we follow
the approach developed in Ref.~\onlinecite{stringari1996} for atomic
condensates in three dimensional traps, which can be directly applied to the
2D case. Rather than considering small deviations of the wave function, let
us now describe the collective excitations of the ground state by the
variation of its density $\overline{\delta c_{TF}(\mathbf{r,}t)}=\delta
c_{TF}(\mathbf{r})=\left[ n(\mathbf{r})-n_{0TF}(\mathbf{r})\right] \mathrm{e}%
^{\mathrm{i}\omega _{TF}t}$ and frequency $\omega _{TF}$. The equation for $%
\delta c_{TF}$ can be derived after some transformation of the GPE including
the linearization procedure and omitting the kinetic energy terms. We then
finally arrive to the following hydrodynamic equation:
\begin{equation}
\omega ^{2}\delta c_{TF}=-\frac{1}{2}\omega _{0}^{2}\nabla \left(
r_{0}^{2}-r^{2}\right) \nabla \delta c_{TF}\:, 
\label{eq:hydro}
\end{equation}%
defined in the same region as Eq.~(\ref{eq:ground}). The solution of (\ref%
{eq:hydro}) can be cast as
\begin{equation}
\delta c_{TF}(\mathbf{r})_{n,m_{z}}=P_{2n,m}(r/r_{0})\left( r/r_{0}\right)
^{m}\mathrm{e}^{\mathrm{i}m_{z}\varphi }\text{ },  \label{eq:delta_rho}
\end{equation}%
where $P_{2n,m}(x)=\sum\limits_{k=0}^{k=n}d_{2k}x^{2k}$ and the coefficients
$d_{2k}$ satisfy the recurrence relation
\begin{equation*}
d_{2k+2}=d_{2k}\frac{(n+m+k+1)(k-n)}{(m+k+1)(k+1)}
\end{equation*}%
for $k=0,1..$ and $d_{0}=1$. The dispersion relation for the excitation modes is
given by the formula (cf. Ref.~(\onlinecite{stringari1996}))%
\begin{equation}
\omega _{TFn,m}=\omega _{0}\sqrt{2n^{2}+2n+2nm+m}\text{ }.  \label{TF}
\end{equation}%
Following the same trends as in the 3D case (see Ref. \onlinecite{pitaevskii}%
) in the limit of strong particle interactions, the excitation energies
become $\Lambda $ independent. This result is represented in Fig.~\ref%
{fig:fig1} by dashed lines. One can see again that the frequencies of the
dipole mode ($n=0$, $m=1$) $\omega _{TFn=0,m=1}=\omega _{0}$ in the Thomas-Fermi
and perturbative approaches indeed coincide. Moreover, the excited state $%
N=2 $, $m_{z}=0$ in both considered limits is also harmonic, i.e. 
$\omega _{N=2,m=0}=\omega _{TFn=1,m=0}=2\omega _{0}.$

\subsection{Probing the excited states of the condensate}

\begin{figure}[tbp]
\includegraphics[width = 0.45\textwidth]{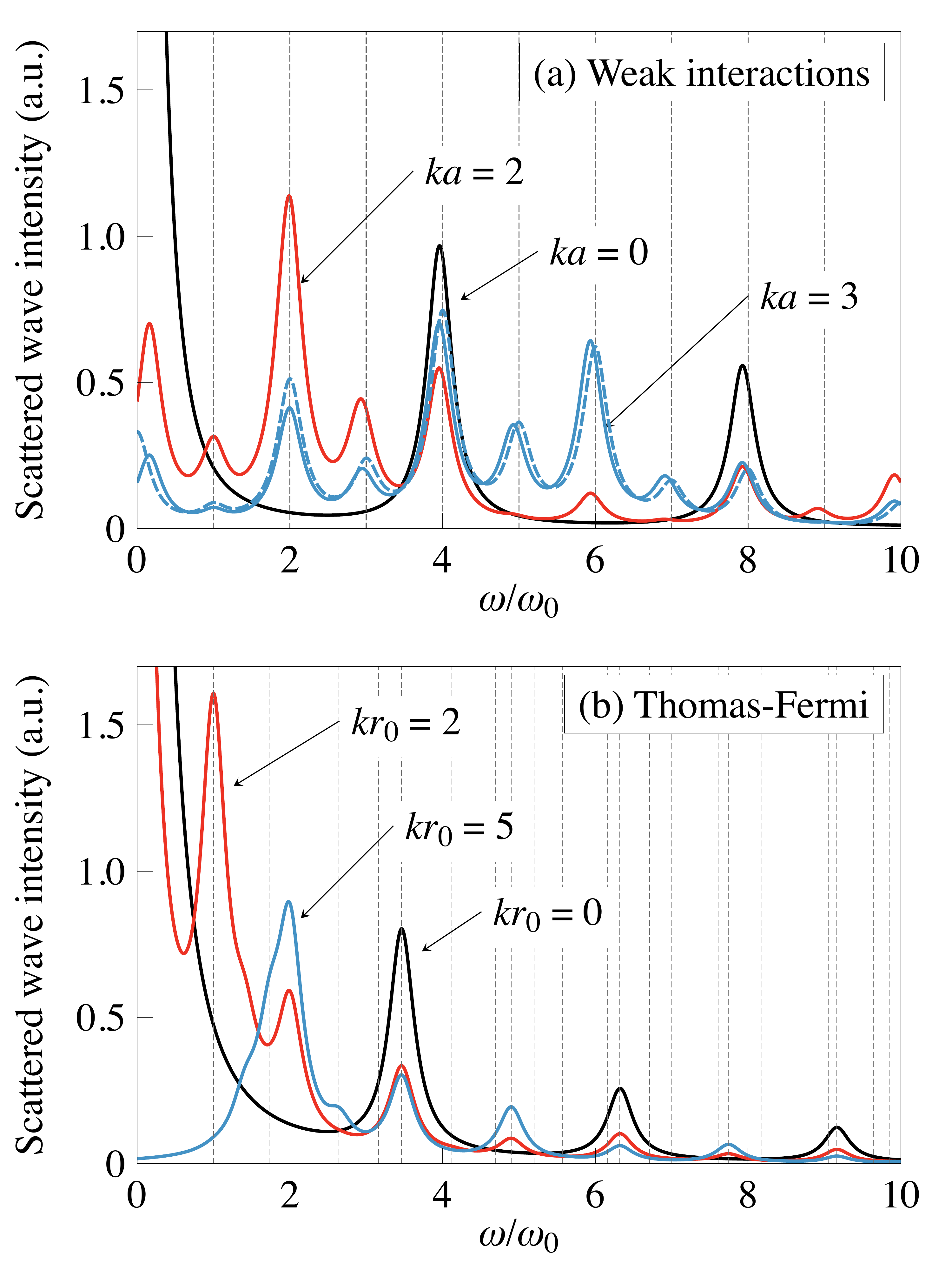}
\caption{(Color online): Spectra of the scattered field in the limit of weak
(a) and strong (b) interactions. Dashed vertical lines indicate the
frequencies calculated by Eq.~\eqref{freque} at $\Lambda =0$ (a) and by Eq.~
\eqref{TF} (b). For comparison, the spectrum at $ka=3$ and $\Lambda =0$ is
shown in panel (a) by the dashed curve.}
\label{fig:fig4}
\end{figure}

Light scattering spectroscopy is a common technique for investigation of the
optical properties of semiconductors,~\cite{spectr1, spectr2, spectr3} which
can be used for probing the excited states of the Bose-Einstein condensates
in microcavities. Consider the two dimensional exciton-photon system in the
strong coupling regime probed by a low-intensity resonant electromagnetic
wave. The scattered wave can be found within the so called input-output
approach~\cite{Collett, Walls_book}. The internal polariton modes are
resonantly excited by the incident field so that Eq.~\eqref{eq:GPT} gains an
additional \textquotedblleft pump\textquotedblright\ term and can be written
as~\cite{Collett}
\begin{equation}
i\hbar \partial _{t}\Psi =\left( -\frac{\hbar ^{2}}{2m_{\ast }}\Delta +V(%
\mathbf{r})+\lambda \left\vert \Psi \right\vert ^{2}\right) \Psi
+\gamma _{d}t_{A}E(\bm{r},t)\text{ }.  \label{eq:resonant}
\end{equation}%
Here $E(\bm{r},t)$ is the incident field, which can be written as $E=E_{0}%
\mathrm{e}^{\mathrm{i}\bm{k}\bm{r}}\mathrm{e}^{-\mathrm{i}\mu t/\hbar }\cos
\omega t$, $\bm{k}$ is the in-plane wave vector ($\bm{k}=0$ under normal
incidence), $\gamma _{d}$ is the cavity damping parameter, and $t_{A}$ is the amplitude transition coefficient of the cavity
mirrors. The wave function $\Psi $ should be treated as the polarization of
an exciton mode or the electric field of a photon mode. Substituting in Eq.~ %
\eqref{eq:resonant} the wave function in the form~(\ref{eq:wafeper}) yields
the problem identical to Eq.~(\ref{eq:bogoliubov}) but with a nonzero right
part $\left[ -1/2tE_{0}\mathrm{e}^{\mathrm{i}\bm{k}\bm{r}}:1/2tE_{0}\mathrm{e%
}^{\mathrm{i}\bm{k}\bm{r}}\right] ^{T}$. Using the Green's function
formalism~\eqref{eq:Green}, the solution of this non-uniform problem is readily obtained. Therefore the output (scattered) field in the positive
frequency range can be written as
\begin{multline}
E_{\mathrm{out}}(\bm{r})\propto t_{A}^{\ast }\Psi =\frac{1}{2}%
|t_{A}|^{2}\gamma _{d}E_{0}\mathrm{e}^{-\mathrm{i}\dfrac{\mu }{\hbar }%
t}\times  \label{eq:outfield} \\
\sum_{N,m}\frac{\varkappa _{N,m_{z}}u_{N,m_{z}}(r)\mathrm{e}^{\mathrm{i}%
m_{z}\theta }}{\hbar \left( \omega _{N,m}-\mathrm{i}\gamma _{d}-\omega
\right) }\mathrm{e}^{-\mathrm{i}\omega _{N,m}t}\text{ },
\end{multline}%
\begin{equation*}
\varkappa _{N,m}(\bm{k})=\frac{1}{2\pi }\int \left[ u_{N,m_{z}}^{\ast
}-v_{N,m_{z}}^{\ast }\right] \mathrm{e}^{-\mathrm{i}m_{z}\theta ^{\prime }}%
\mathrm{e}^{\mathrm{i}\bm{k}\bm{r}^{\prime }}d\bm{r}^{\prime }\text{ }.
\end{equation*}%
A possible experimental observable is the integrated intensity of the
scattered wave given by
\begin{equation}
\frac{4\int \left\vert E_{\mathrm{out}}(\bm{r})\right\vert ^{2}d\bm{r}}{%
|t_{A}|^{4}E_{0}^{2}}=\gamma _{d}^{2}\sum_{N,m_{z}}\frac{|\varkappa
_{N,m_{z}}|^{2}}{\hbar ^{2}\left[ (\omega _{N,m}-\omega )^{2}+\gamma _{d}^{2}%
\right] }\text{ }.  \label{eq:scattered}
\end{equation}%
This quantity is represented in Fig.~\ref{fig:fig4}(a) for different values
of the wave vector $\mathbf{k}$ (corresponding to different angles of
incidence of the probing wave). One can see that at normal incidence ($k=0$)
only the modes with $m=0$ are excited and due to parity considerations the
modes with even radial numbers $n$ exhibit considerably stronger interaction
with the probing field. At $k\neq 0$ the modes with non-zero $m$ appear and
the peak value shifts towards higher energies with increasing $k$.

In the Thomas-Fermi limit, the expression for the field analogous to Eq.~%
\eqref{eq:outfield} is obtained if one takes the wave functions of
excitations in the form of $\delta c_{TF}(\mathbf{r})_{n,m_{z}}/\sqrt{%
n_{0TF}(\bm{r})}$. The scattered spectrum then has the same form as given by
Eq.~\eqref{eq:scattered} with the following $\varkappa _{N,m_{z}}$
coefficients:
\begin{equation}
\varkappa _{n,m_{z}}(\bm{k})=\int \frac{\delta c_{TF}(\bm{r}^{\prime
})_{n,m_{z}}}{\sqrt{c_{0TF}(\bm{r}^{\prime })}}\mathrm{e}^{-\mathrm{i}%
m_{z}\theta ^{\prime }}\mathrm{e}^{\mathrm{i}\bm{k}\bm{r}^{\prime }}d\bm{r}%
^{\prime }\text{ }.
\end{equation}%
The calculated spectra are presented in Fig.~\ref{fig:fig4}(b).

\section{Quasi one-dimensional exciton-polariton condensates}

In the following, we reduce the two dimensional GPE (\ref{eq:GPT}) to a 1D
problem along the axial direction ($x$) of a microwire. This can be done by
"freezing out" the $y-$motion of the condensate (due to the presence of a
lateral confinement potential) and by re-normalizing the mean-field
interaction. We consider a wire cavity with a separable potential $V(\mathbf{%
r})=U_{x}(x)+U_{y}(y)$, where $U_{x}(x)=\frac{1}{2}m\omega _{0y}x^{2}$ is the
harmonic trap and $U_{y}$ is the cavity confinement potential along $y-$%
axis. Employing the adiabatic approximation, the order parameter $\Psi (%
\mathbf{r},t)$ can be factorized as

\begin{equation}
\Psi (\mathbf{r},t)=\sqrt{\mathcal{N}}\phi _{x}(x,t)\exp
(-iE_{_{y}}t/\hslash )\phi _{y}(y)\text{ },  \label{Fact}
\end{equation}%
where the longitudinal wave function $\phi _{y}$ and the energy $E_{_{y}}$
are determined by the auxiliary problem

\begin{equation*}
E_{_{y}}\phi _{y}=\left[ \frac{p_{y}^{2}}{2m_{\ast }}+U_{y}\right] \phi _{y}%
\text{ }.
\end{equation*}%
Assuming an infinite confinement barrier, the polariton eigenenergies are $%
E_{_{y}}=\hslash \omega _{n_{y}}$ with $\omega _{n_{y}}=\dfrac{\hslash }{2m}%
\left( \dfrac{n_{y}\pi }{L_{y}}\right) ^{2},$ $n_{y}$ is an integer number, and
$L_{y}$ denotes the microwire cavity length. (Notice that the adiabatic
approximation is restricted to the strong spatial confinement case where the
frequency $\omega _{0x}\ll \omega _{n_{y}=1}).$ Substituting Eq.~(\ref{Fact}%
) into Eq.~(\ref{eq:GPT}) and assuming that the condensate along the $y-$%
direction remains close to the $\omega _{n_{y}=1}$ ground state with $\phi
_{n_{y}=1}=\sqrt{2/L_{y}}\sin (\pi x/L_{y}),$ we obtain the equation~\cite{Nota1} 

\begin{equation}
i\hbar \partial _{t}\phi _{x}=\left( \frac{p_{x}^{2}}{2m_{\ast }}%
+U_{x}+g_{1}\left\vert \phi _{x}\right\vert ^{2}\right) \phi _{x}\text{ },
\label{1DGPE}
\end{equation}%
where $g_{1}=3\lambda \mathcal{N}/(2L_{y})$ is the effective 1D
polariton-polariton interaction constant$.$~\cite{Nota}

\subsection{Normal modes in 1D-parabolic potential}

In this case, we employ the same method of the linear response as in
Section II with $V(\mathbf{r})\rightarrow m_{\ast }\omega _{0x}^{2}x^{2}/2,$
$\Lambda \rightarrow \Lambda _{1D}=g_{1}/(l_{x}\hslash \omega _{0x}),$ and $%
l_{x}=\sqrt{\hslash /(m_{\ast }\omega _{0x})}.$ The present problem is
analogous to the one studied for diluted atomic gases in 1D optical lattices
(see Refs.~\onlinecite{trallero1} and \onlinecite{trallero2}). Applying
those results to our case we have for the 1D chemical potential:

\begin{equation}
\mu _{1D}=\omega _{0x}\left( \frac{1}{2}+\frac{\Lambda _{1D}}{\sqrt{2\pi }}%
+C_{1}\Lambda _{1D}^{2}\right)  \label{chepo1D}
\end{equation}%
and for the 1D excitation frequencies,

\begin{multline}
\omega _{k}^{(1D)}=\omega _{0x}\left[ k+\frac{\Lambda _{1D}}{\sqrt{2\pi }}%
\left( -1+\frac{2\Gamma (k+1/2)}{\sqrt{\pi }k!}\right) \right.
\label{frecue1D} \\
\left. \Lambda _{1D}^{2}\left( \frac{\gamma _{k}}{2\pi ^{2}}-C_{1}\right) %
\right] \text{ };\text{ \ \ \ \ \ }k=1,2,....\text{ ,}
\end{multline}%
where $C_{1}=\frac{3}{2\pi }\ln \left[ \frac{\sqrt{3}}{4}+\frac{1}{2}\right]
$, $\Gamma (z)$ is the gamma function and $\gamma _{k}$ are numeric
parameters given elsewhere.~\cite{trallero2} The corresponding total density
in the excited state $k,$ $c_{k}^{(1D)}(x,t),$ and the excitation profile, $%
\delta c_{k}^{(1D)}(x,t)$ are displayed in the Appendix B.

\begin{figure}[tbp]
\includegraphics[width = 0.48\textwidth]{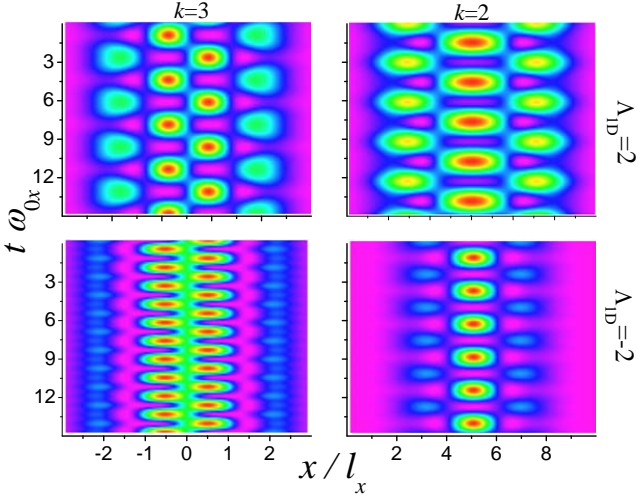}
\caption{(Color online): Evolution of the 1D condensate density, $%
c_{k}^{(1D)}(x,t)$, for the excited state $\protect\omega _{k}^{(1D)}$ ($k=2$
and 3) and 1D self-interaction parameter $\Lambda _{1D}=\pm 2$. }
\label{fig:fig5}
\end{figure}
The dynamics of the condensate calculated using Eqs.~(\ref{C1D})-(\ref%
{delC1D}) are sketched in Fig.~\ref{fig:fig5} by a 2D map of the density, $%
c_{k}^{(1D)}(x,t)$ ($k=2$ and 3) as function of dimensionless coordinate $%
x/l_{x}$ and time $t\omega _{0x}.$ For the calculation we chose $\Lambda
_{1D}=\pm 2.$ From the figure we observe that, for a certain moment of time,
there are pronounced oscillations of the density, $c_{k}^{(1D)}(x,t),$ along
the $x-$axis quenched according to the exponential behavior $%
exp(-x^{2}/l_{x}^{2}).$ Moreover, from Fig.~\ref{fig:fig5} it becomes clear
that the condensate is stronger localized in space in the case of attractive
polariton-polariton interaction (negative sign of $\Lambda _{1D})$.

\subsection{Normal modes in 1D semi-parabolic potential}

Let us now consider a semi-parabolic potential~\cite{Wertz,Ferrier}
\begin{figure}[tbp]
\includegraphics[width = 0.40\textwidth]{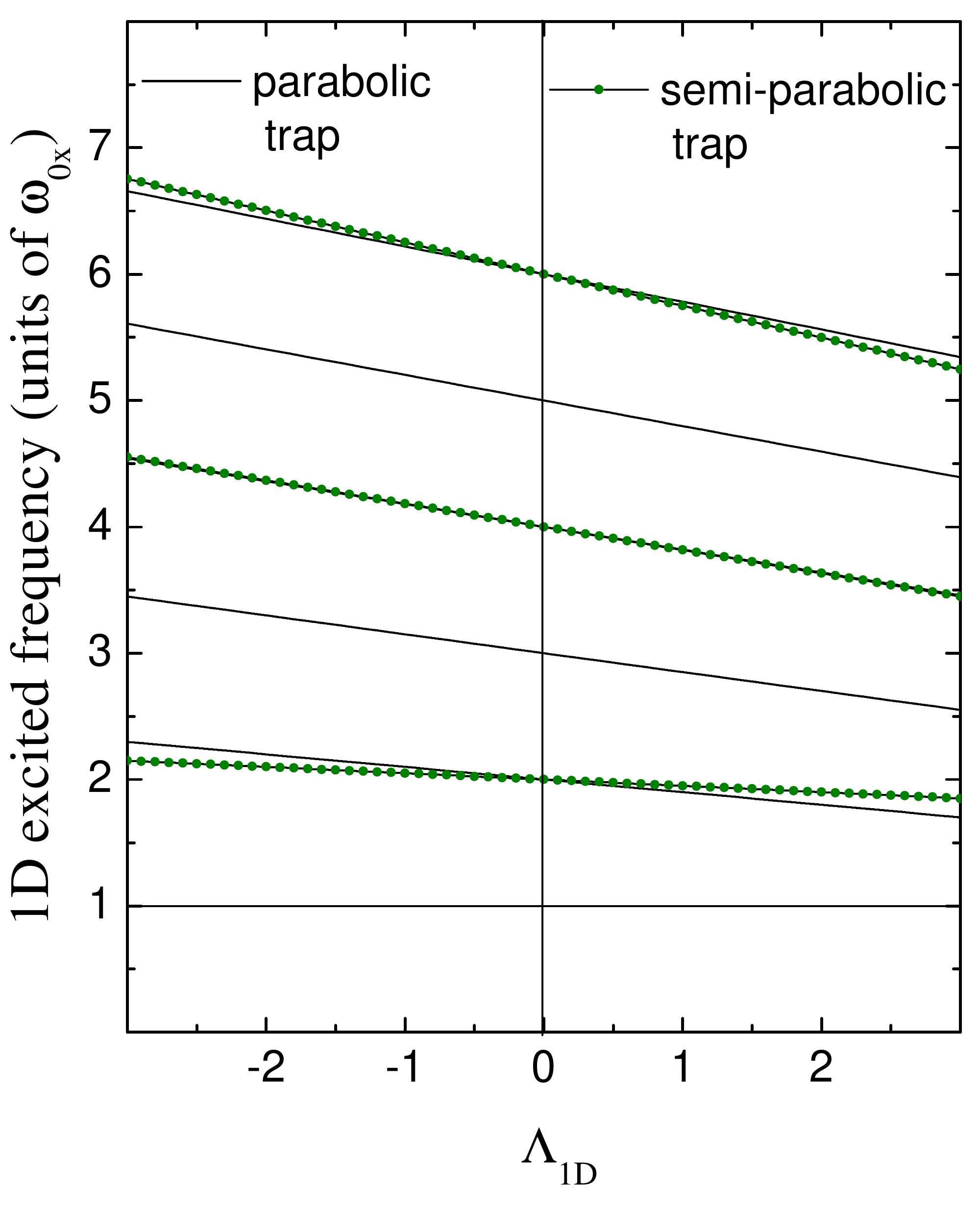}
\caption{(Color online): Collective excitation frequencies $\protect\omega %
_{k}^{(1D)}$ (solid lines) and $\protect\omega _{k}^{(1/2D)}$ (dot-solid
lines) for 1D parabolic and semi-parabolic trap potentials, respectively, as
a function of 1D dimensionless parameters $\Lambda _{1D}$. Note that $%
\protect\omega _{1}^{(1D)}$ has the frequency value $\protect\omega _{0}$ of
the harmonic trap (see Ref.~\onlinecite{pitaevskii}).}
\label{fig:fig6}
\end{figure}
\begin{equation}
V(x)=\left\{
\begin{array}{c}
0\text{ \ \ \ \ \ \ \ \ \ ; \ \ \ \ }x<0\text{ } \\
\frac{1}{2}m\omega _{0x}^{2}x^{2}\text{ ; \ }0<x<\infty%
\end{array}%
\right. \text{ .}  \label{parabolic}
\end{equation}%
In this case the order parameter must fulfill the boundary conditions $\psi
_{0}^{(1/2D)}(x=0)=0$ and $\psi _{0}^{(1/2D)}\rightarrow 0$ at $x\rightarrow
\infty $. The solution of GPE (\ref{1DGPE}) with the semi-parabolic
potential (\ref{parabolic}) yields the following expression for the chemical
potential $\mu _{1/2D}$ obtained up to second order in $\Lambda _{1D}$ (see
Appendix C)

\begin{equation}
\mu _{1/2D}=\omega _{0x}\left( \frac{3}{2}+\frac{3}{2\sqrt{2\pi }}\Lambda
_{1D}+C_{1/2}\Lambda _{1D}^{2}\right) \text{ .}
\end{equation}%
The corresponding Bogolyubov's excitaion frequencies{\large \ }$\omega
_{k}^{(1/2D)}${\large \ }can be cast as

\begin{multline}
\omega _{k}^{(1/2D)}=\omega _{0x}\left[ 2k+\frac{\Lambda _{1D}}{\sqrt{2\pi }}%
\left( -\frac{3}{2}+\frac{4(2k+3/4)(4k)!}{(2k+1)!2^{4k}(2k)!}\right) \right.
\label{parabolic1/2} \\
\left. \Lambda _{1D}^{2}\left( \frac{\gamma _{k}^{\prime }}{\pi ^{2}}%
-C_{1/2}\right) \right] \text{ };\text{ \ \ \ \ \ }k=1,2,....\text{ ,}
\end{multline}%
with $\gamma _{k}^{\prime }$ numeric parameters obtained in Appendix C.

Using the above analytical solutions, the frequencies $\omega _{k}^{(1D)}$ and $%
\omega _{k}^{(1/2D)}$\ for the first 6 and 3 modes, respectively, {\it versus} the self-interaction parameter $\Lambda _{1D}$ (for attractive, $%
\Lambda _{1D}<0$ and repulsive $\Lambda _{1D}>0$ polariton-polariton
interactions) are represented in Fig.~\ref{fig:fig6}. The
symmetry of the semi-parabolic trap requires that only odd states exist (see
Appendix C). It means that the first excited state corresponds to $\omega
_{k=1}^{(1/2D)}$ and, unlike the case of harmonic potential, its energy
depends on $\Lambda _{1D}$. This is a consequence of the fact that potential
(\ref{parabolic}) breaks the inversion symmetry and the Konn's theorem is
not valid.

\section{Conclusions}

In summary, we have studied the two- and one-dimensional Bogolyubov's
excitation modes of a Bose-Einstein condensate of exciton-polaritons in 2D
in microcavities and microwire-cavities with harmonic traps. In the 2D case
we have found eigenenergies and eigenfunctions of the collective modes for
two limiting regimes: the weak and strong polariton-polariton interactions.
In the weak interaction limita and based on the perturbative method of solution of the non-linear GPE,
 we derived explicit analytical
expressions for the collective excitation frequencies given by Eqs.~(\ref%
{freque}), (\ref{frecue1D}), and (\ref{parabolic1/2}). In the
two-dimensional case, there are two independent spaces of solutions, $%
\mathcal{I}$ and $\mathcal{II}$ (see Appendix A), and the corresponding
excitation spectrum is ruled by the angular momentum conservation. In all
considered cases, the Bogolyubov's frequencies plotted against the 
self-interaction parameter, $\lambda ,$ show a negative slope. In the case
of strong polariton-polariton interaction case, where the Thomas-Fermi
approximation is valid, the wave functions and the eigenfrequencies are
presented by Eqs.~(\ref{eq:delta_rho}) and (\ref{TF}) with the dispersion law $%
\lambda $ independent.

We have shown that in 1D traps the polariton-polariton coupling strength is
renormalised, and it can be controllued by the confinement potential of the
frozen $y-$motion (and scales as $1/L_{y}$), allowing for a modulation of the
non-linear cubic term, $g_{1}\left\vert \phi _{x}\right\vert ^{3},$ and,
consequently, it affects the spectrum of the excited states. Also, we derived the
complete sets of the 2D and 1D excitation modes, which allow for the
calculation of a variety of dynamical variables relevant to experiments. In
particular, we presented a theory on light scattering by the confined microcavity condensate and calculated, in
both considered limits, the spectral dependence of the integrated intensity
of a scattered electromagnetic wave. We have
calculated the polariton current density associated with the elementary
excitations. It is related to the density profile for the excited states
(Figs.~\ref{fig:fig2} and \ref{fig:fig3}). We suggest that they are relevant
to the experimentally measured real-space spectra distribution for the
polariton pendulum (see Figs.~1 and 2 in Refs.~%
\onlinecite{tosi,KavovinPolariton}) and spatially mapped exciton-polariton
condensate wave functions (see Fig.~4 in Ref.~\onlinecite{Wertz}) are quite
well reproduced by the density profiles $c_{k}^{(1D)}$ and $c_{k}^{(1/2D)}$ (%
$k=1,2...$) given in the Appendices B and C. A good agreement of theory
and experiment is found for small values of the $g_{1}$ parameter, which
corresponds to the experimental setting of Ref.~\onlinecite{Wertz}.

\section*{Acknowledgments}

C.T.-G. and V.L-R. acknowledge the financial support of Brazilian agencies
CNPq and FAPESP. M.V. acknowledges support from the Portuguese Foundation for Science and
Technology through Projects PTDC-FIS-113199-2009 and PEst-C/FIS/UI0607/2013. M.D. is grateful to M.
Glazov for valuable discussions and acknowledges the financial support from
the Dynasty Foundation, RFBR, and EU projects POLAPHEN and SPANGL4Q. M.D.
and A.K. acknowledge Russian Ministry of Education and Science (Contract No.
11.G34.31.0067 with SPbSU).

\appendix

\section{2D eigenmodes}

The radial functions for the 2D harmonic oscillator are: $R_{N,m}=e^{-\frac{\rho ^{2}%
}{2}}\rho ^{m}L_{n}^{(m)}(\rho ^{2})/\sqrt{\mathcal{N}_{N,m}},$, where $%
L_{n}^{(m)}(z)\mathcal{\ }$ are the Generalized Laguerre polynomials~\cite%
{Abramowitz}, $n=(N-m)/2=0,1,...$ is the radial number, and $\mathcal{N}_{N,m}=(%
\frac{N+m}{2})!/\left( \frac{N+m}{2}\right) !$ is a normalization constant.%
{\Large \ }Based on the symmetry consideration, the solutions of Eqs.~(\ref%
{eq:bogoliubov}) can be classified accordingly to the parity of the
z-component of the angular momentum, i.e. we have two independent space of
solutions; $\left\vert u_{N,m}^{(I)}\right\rangle $\ and $\left\vert
^{(I)}v_{N,m}\right\rangle $\ for $m$ even and $\left\vert
u_{N,m}^{(II)}\right\rangle $\ and $\left\vert ^{(II)}v_{N,m}\right\rangle $
with $m$ odd. After substitution of (\ref{serie1}) into (\ref{eq:bogoliubov}%
), the eigenvalue problem is reduced to the system of linear equations:
\begin{eqnarray}
&&\left. \Lambda \sum\limits_{N_{1}}\langle R_{N_{2},m}|\overline{n_{0}}%
|R_{N_{1},m}\rangle \left( 2A_{NN_{1}}+B_{NN_{1}}\right) \right.   \notag \\
&=&(\varpi _{Nm}-N_{2}-1+\mu /\omega _{0})A_{NN_{2}}\text{ },
\label{eigenvalueA}
\end{eqnarray}%
\begin{eqnarray}
&&\left. -\Lambda \sum\limits_{N_{1}}\langle R_{N_{2},m}|\overline{n_{0}}%
|R_{N_{1},m}\rangle \left( 2B_{NN_{1}}+A_{NN_{1}}\right) \right.   \notag \\
&=&(\varpi _{Nm}+N_{2}+1-\mu /\omega _{0})B_{NN_{2}}\text{ },
\label{eigenvalueB}
\end{eqnarray}%
where $\varpi _{N,m}=\omega _{N,m}/\omega _{0}$ are the dimensionless
Bogolyubov frequencies. The reduced frequencies $\varpi _{Nm}$, the
coefficients $A_{NN_{1}}$ and $B_{NN_{1}}$ can be written in a form of
Taylor expansions:%
\begin{eqnarray}
\varpi _{N,m} &=&\sum_{i=0}^{\infty }\varpi _{N,m}^{(i)}\Lambda ^{i}\text{ },
\notag \\
A_{NN_{1}}\left[ B_{NN_{1}}\right]  &=&\sum_{i=0}^{\infty }A_{NN_{1}}^{(i)}%
\left[ B_{NN_{1}}^{(i)}\right] \Lambda ^{i}\text{ }.  \label{expantion}
\end{eqnarray}%
\bigskip Using the series (\ref{expantion}) and Eqs.~(\ref{eigenvalueA}) and
(\ref{eigenvalueB}) at \textit{\ zeroth order} in $\Lambda $ we get

\begin{equation}
\varpi _{N,m}^{(0)}=N\mbox{; }B_{NN_{2}}^{(0)}=0\mbox{; }A_{NN_{2}}^{(0)}=%
\delta _{NN_{2}}\text{ }.  \label{zero}
\end{equation}%
Taking the \textit{first order terms }in Eq.~(\ref{eigenvalueA}) we have
\begin{eqnarray}
2\left\langle N_{2},m\left\vert n^{(0)}\right\vert N,m\right\rangle
&=&\left( \varpi _{N,m}^{(1)}+\frac{1}{2\pi }\right) \delta _{N,N_{2}}
\notag \\
&&+\left( N-N_{2}\right) A_{NN_{2},m}^{(1)}\text{ .}
\label{matricial1 first}
\end{eqnarray}%
Using the condensate distribution density (\ref{eq:groundstate_psi1}) we can
write that
\begin{eqnarray*}
&&\left. \left\langle N_{2},m\left\vert n^{(0)}\right\vert N,m\right\rangle
\equiv \right. C_{N,N_{2}}^{(0)}= \\
&&\frac{1}{2\pi \mathcal{N}_{N,m}}\int_{0}^{\infty
}L_{n}^{(m)}(t)L_{n_{2}}^{(m)}(t)^{2}\times t^{m}\exp (-2t)dt\text{ .}
\end{eqnarray*}%
This integral can be calculated in quadrature:~\cite{GRTable}
\begin{equation}
C_{N,N_{2}}^{(0)}=\frac{2^{-n-n_{2}-m-1}(n+n_{2}+m)!}{\pi \sqrt{%
(n+m)!(n_{2}+m)!n!n_{2}!}}\text{ .}  \label{A6}
\end{equation}%
Equation (\ref{matricial1 first}) for $N_{2}=N$ yields:
\begin{equation}
\varpi _{N,m}^{(1)}=\frac{1}{2\pi }\left\{ 2^{-N+1}\left(
\begin{array}{c}
N \\
(N-m)/2%
\end{array}%
\right) -1\right\} \mbox{,}  \label{lambda1}
\end{equation}%
and for $N_{2}\neq N$, it leads to:%
\begin{equation}
A_{NN_{2},m}^{(1)}=\frac{2}{N-N_{2}}C_{N,N_{2},m}^{(0)}\text{ ,}
\label{A1 1}
\end{equation}%
while $A_{NN,m}^{(1)}=0$ from normalization. In the same way, from (\ref%
{eigenvalueB}), (\ref{zero}) and (\ref{matricial1 first}), we obtain
\begin{equation}
B_{NN_{2},m}^{(1)}=-\frac{1}{N+N_{2}}C_{N,N_{2},m}^{(0)}\text{ .}  \label{B1}
\end{equation}%
Accordingly, collecting \emph{second order terms} in $\Lambda $ from (\ref%
{eigenvalueB}) we have,
\begin{eqnarray}
&&\left. \sum_{N_{1}}\left\langle N_{2},m\left\vert n^{(0)}\right\vert
N_{1},m\right\rangle \left[ 2A_{NN_{1},m}^{(1)}+B_{NN_{1},m}^{(1)}\right]
\right.  \notag \\
&&\left. +2\left\langle N_{2},m\left\vert n^{(1)}\right\vert
N,m\right\rangle =\left( N-N_{2}\right) A_{NN_{2},m}^{(2)}+\right.  \notag \\
&&\left. \left( \lambda _{N,m}^{(1)}+E^{(1)}\right)
A_{NN_{2},m}^{(1)}+\left( \lambda _{N,m}^{(2)}+E^{(2)}\right) \delta
_{NN_{2}}\right. \text{ .}  \notag \\
&&  \label{2nd}
\end{eqnarray}%
If $N_{2}=N$, Eq.~(\ref{2nd}) reads:

\begin{equation}
\varpi _{N,m}^{(2)}=S_{N,m}+2C_{N,m}^{(1)}-\frac{1}{2N}\left(
C_{N,m}^{(0)}\right) ^{2}+\frac{3}{8\pi ^{2}}\ln (\frac{4}{3})\text{ ,}
\label{lambda2}
\end{equation}%
where
\begin{equation}
S_{N,m}=\sum_{N_{1}\neq N}\left[ \left( C_{N,N_{1},m}^{(0)}\right) ^{2}\frac{%
3N+5N_{1}}{(N-N_{1})(N+N_{1})}\right] \mbox{,}
\end{equation}%
\begin{equation}
C_{N,m}^{(1)}=\frac{1}{2\pi \mathcal{N}_{N,m}}\int_{0}^{\infty }\left(
L_{n}^{(m)}(t)\right) ^{2}t^{m}F(t)\exp (-2t)dt\text{ .}  \label{coefi}
\end{equation}%
Using the results~\cite{GRTable,Abramowitz}%
\begin{equation}
\int_{0}^{\infty }t^{a}\exp (-2t)dt=2^{-a-1}a!\text{ ,}
\end{equation}%
\begin{equation}
\int_{0}^{\infty }\Gamma (0,t)t^{a}\exp (-2t)dt=2^{-a-1}a!B_{2/3}(a+1,0)%
\text{ ,}
\end{equation}%
\begin{equation}
\int_{0}^{\infty }t^{a}\ln t\exp (-2t)dt=-2^{-a-1}a!\left( \gamma -H_{a}+\ln
2\right) \text{ ,}
\end{equation}%
with $H_{a}$ the $a$-th harmonic number, and $B_{x}(a,b)$ the Incomplete
Beta function, we obtain%
\begin{eqnarray*}
I_{a} &=&\int_{0}^{\infty }t^{a}F\exp (-2t)dt= \\
&&\frac{a!}{2^{a+2}\pi }\left\{ -\ln 4+H_{a}+B_{{2}/{3}}\left( a+1,0\right)
\right\} \text{ .}
\end{eqnarray*}%
Expanding the Laguerre polynomials as Taylor series~\cite{Abramowitz} follows%
\begin{equation*}
\left( L_{n}^{(m)}(t)\right) ^{2}=\sum_{k,\text{ }l=0}^{n}\frac{(-1)^{k+l}}{%
k!l!}\left(
\begin{array}{c}
n+m \\
n-k%
\end{array}%
\right) \left(
\begin{array}{c}
n+m \\
n-l%
\end{array}%
\right) t^{k+l}
\end{equation*}%
and inserting in Eq.~(\ref{coefi}) we have
\begin{eqnarray*}
C_{N,m}^{(1)} &=&\frac{1}{\pi }\frac{n!}{(n+m)!}\sum_{k=0}^{n}\sum_{l=0}^{n}%
\frac{(-1)^{k+l}}{k!l!}\times \\
&&\left(
\begin{array}{c}
n+m \\
n-k%
\end{array}%
\right) \left(
\begin{array}{c}
n+m \\
n-l%
\end{array}%
\right) I_{m+k+l}\text{ .}
\end{eqnarray*}%
For the functions $u_{N,m}$ and $v_{N,m}$ up to first order in $\Lambda $ we
obtain:
\begin{eqnarray}
u_{N,m} &=&R_{N,m}+2\Lambda \sum_{N_{2}\neq N}\frac{%
C_{N,N_{2}}^{(0)}R_{N_{2},m}}{N-N_{2}}\text{ }, \\
v_{N,m} &=&\Lambda \sum_{N_{2}}\frac{C_{N,N_{2}}^{(0)}R_{N_{2},m}}{N+N_{2}}%
\text{ }.
\end{eqnarray}

\section{1D eigenmodes for parabolic potential}

The concentration $c_{k}^{(1D)}$ in the excited state $k$ is given by

\begin{equation}
c_{k}^{(1D)}=n_{0}^{(1D)}(x/l_{x})+\delta c_{k}^{(1D)}(x/l_{x},t)\text{ ,}
\label{C1D}
\end{equation}%
where~\cite{trallero1}

\begin{equation}
n_{0}^{(1D)}(z)=\frac{1}{\sqrt{\pi }}\exp (-z^{2})+\Lambda _{1D}\sqrt{\frac{2%
}{\pi ^{3/2}}}\exp (-z^{2})\mathcal{F}(z)\text{ ,}  \label{n1D}
\end{equation}%
\begin{equation*}
\mathcal{F}(z)=\int\limits_{1}^{\sqrt{2}/2}\dfrac{\exp (-\frac{z^{2}}{y^{2}}%
(1-y^{2}))-1}{1-y^{2}}dy\text{ ,}
\end{equation*}%
\begin{multline}
\delta c_{k}^{(1D)}(z,t)=2\cos \left( \omega _{k}^{(1D)}t\right) \left(
\sqrt{n_{0}^{(1D)}}\varphi _{k}(z)\right.  \label{delC1D} \\
-2\Lambda _{1D}\varphi _{0}(z)\left\{ \sum\limits_{m\neq k\text{ };\text{ }%
m\neq 0}\left[ \frac{1}{m-k}+\frac{1}{2(m+k)}\right] \right. \\
\left. \left. \times T_{00mk}\varphi _{m}(z)+\frac{1}{2^{2}k}T_{00kk}\varphi
_{k}(z)\right\} \right) \text{ ,}
\end{multline}%
and

\begin{equation*}
T_{00mk}=\frac{\left( -1\right) ^{\frac{k-m}{2}}}{\pi \sqrt{2m!k!}}\Gamma
\left( \frac{m+k+1}{2}\right) \text{ .}
\end{equation*}%
Here $\varphi _{k}(z)$ is the harmonic oscillator function

\begin{equation}
\varphi _{k}(z)=\frac{1}{\sqrt{\sqrt{\pi }2^{k}k!}}\exp (-\frac{z^{2}}{2}%
)H_{k}(z)\text{ },\text{ \ \ \ \ \ }k=0,1,2...  \label{osc}
\end{equation}%
with $H_{l}(x)$ denoting the Hermitian polynomial.~\cite{Abramowitz}

\section{1D eigenmodes for semi-parabolic potential}

Considering the potential (\ref{parabolic}), the solution of the
one-dimensional nonlinear GPE can be sought in terms of the complete set of
functions \{$\varphi _{k}^{(1/2)}(x/l_{x})=\sqrt{2}\varphi _{2k+1}(x/l_{x})$%
\}. Taking only interaction terms up to second order in $g_{1}$, we obtain
for the chemical potential%
\begin{equation}
\mu _{1/2D}=\omega _{0x}\left[ \frac{3}{2}+\Lambda _{1/2D}\overline{T_{0000}}%
-3\Lambda _{1D}^{2}\sum\limits_{p\neq 0}^{\infty }\dfrac{\overline{T_{000p}}%
^{2}}{2p}\right] \text{ },  \label{chemserie}
\end{equation}%
where

\begin{equation}
\overline{T_{mlkp}}=\int\limits_{0}^{\infty }\varphi _{m}^{(1/2)}\varphi
_{l}^{(1/2)}\varphi _{k}^{(1/2)}\varphi _{p}^{(1/2)}dx\text{ }.
\label{Tmpkl}
\end{equation}%
From (\ref{Tmpkl}) we have that $\overline{T_{0000}}=3/(2\sqrt{2\pi })$ and

\begin{equation}
\overline{T_{000p}}=\sqrt{\frac{2}{\pi }}\frac{\left( -1\right) ^{p+1}}{%
2^{2p}}\frac{\sqrt{(2p+1)!}}{p!}\left( \frac{p}{2}-\frac{3}{4}\right) \text{
.}  \label{Tooop}
\end{equation}%
Following (\ref{Tooop}), the series in Eq.~(\ref{chemserie}) can be summed up

\begin{equation*}
\sum\limits_{p\neq 0}^{\infty }\dfrac{\overline{T_{000p}}^{2}}{2p}=\frac{1}{%
4\pi }\left[ -\frac{3}{2}+\frac{7\sqrt{3}}{9}+9\ln \left( 2\sqrt{(2-\sqrt{3})%
}\right) \right] \text{ .}
\end{equation*}%
For the concentration $n_{0}^{(1/2D)}(x/l_{x})=\overline{n_{0}^{(1/2D)}}%
/l_{x}$ we have

\begin{eqnarray}
\overline{n_{0}^{(1/2D)}}(z) &=&\frac{1}{\sqrt{\pi }}\exp (-z^{2})\left[
H_{1}^{2}(z)-\Lambda _{1D}\sqrt{\frac{2}{\pi }}\times \right.  \notag \\
&&\left. H_{1}(z)\sum\limits_{p\neq 0}^{\infty }\dfrac{\left( -1\right)
^{p+1}}{p2^{3p+1}p!}\left( \frac{p}{2}-\frac{3}{4}\right) H_{2p+1}(z)\right]
\text{ .}  \notag \\
&&
\end{eqnarray}%
Assuming the Bogolyubov's method (\ref{eq:wafeper}) and employing the
expansion $u_{1/2D}(x)[v_{1/2D}(x)]=\sum\limits_{k,i=0}^{\infty
}A_{k}^{(i)(1/2D)}\Lambda _{1/2D}^{i}\left[ B_{k}^{(i)(12D)}\Lambda
_{1/2D}^{i}\right] \varphi _{k}^{(1/2)}(x/l_{x})$ and $\varpi
^{(1/2D)}=\omega _{k}^{(1/2D)}/\omega _{0x}=\sum\limits_{i=0}^{\infty
}\varpi ^{(i)(1/2D)}\Lambda _{1D}^{i}$, the collective excitations are
described by the linear system equations

\begin{multline}
\sum\limits_{i=0}^{\infty }\Lambda _{1/2D}^{i}\sum\limits_{k_{1}}\langle
\varphi _{k}^{(1/2)}|\overline{n_{0}^{(1/2D)}}|\varphi
_{k_{1}}^{(1/2)}\rangle \left( 2A_{k_{1}}^{(i)(1/2D)}+\right.
\label{1DbogoA} \\
\left. B_{k_{1}}^{(i)(1/2D)}\right) =\left( \sum\limits_{i=0}^{\infty
}\varpi ^{(i)(1/2D)}\Lambda _{1D}^{i}-2k\right. \\
\left. -\frac{3}{2}+\frac{\mu _{1/2D}}{\omega _{0x}}\right) \Lambda
_{1D}^{k}A_{k}^{(i)(1/2D)}\text{ },
\end{multline}%
and similar equations but changing $A_{k_{1}}^{(i)(1/2D)}\Leftrightarrow
B_{k_{1}}^{(i)(1/2D)}.$ From these system equations we get at\textit{\
zeroth order }in $\Lambda _{1D}$ the $\varpi _{k}^{(0)(1/2D)}=2k$ and at
\textit{first order},
\begin{equation*}
\varpi _{k}^{(1)(1/2D)}=\Lambda _{1D}\left[ -\frac{3}{2\sqrt{2\pi }}+2%
\overline{T_{00kk}}\right] \text{ ,}
\end{equation*}%
where
\begin{equation*}
\overline{T_{00kk}}=\frac{2\left( 2k+3/4\right) }{\sqrt{2\pi }\left(
2k+1\right) !}\frac{\left( 4k\right) !}{2^{4k}\left( 2k\right) !}\text{ .}
\end{equation*}%
For the second order immediately follows

\begin{multline}
\omega _{k}^{(2)(1/2D)}=-\mu _{1/2D}^{(2)}+4\Lambda _{1D}^{2}\left[ -\sqrt{%
\frac{2}{\pi }}\times \right.  \label{w2} \\
\left. \sum\limits_{p\neq 0}^{\infty }\dfrac{\left( -1\right) ^{p+1}\sqrt{%
(2p+1)!}}{2^{2p}2pp!}\left( \frac{p}{2}-\frac{3}{4}\right) \overline{T_{0pkk}%
}\right. \\
\left. -\frac{1}{2}\sum\limits_{p\neq 0;p\neq k}^{\infty }\left( \frac{1}{p-k%
}+\frac{1}{4(p+k)}\right) \overline{T_{00pk}}^{2}\right. \\
\left. -\frac{1}{16k}\overline{T_{00kk}}^{2}\right] \text{ };\text{ \ \ \ \
\ }k=1,2,....\text{ ,}
\end{multline}%
with

\begin{eqnarray}
\overline{T_{00kp}} &=&\frac{\left( -1\right) ^{k-p}}{\pi }\frac{\sqrt{2}%
\Gamma \left( k+p+1/2\right) }{\sqrt{\left( 2k+1\right) !\left( 2p+1\right) !%
}}\times  \notag \\
&&\left[ \frac{1}{4}-\left( k-p\right) ^{2}+k+p+\frac{1}{2}\right] \text{ }
\label{Tkp}
\end{eqnarray}%
and

\begin{eqnarray}
\overline{T_{0kkp}} &=&\frac{1}{\pi ^{2}\sqrt{2}}\frac{\Gamma ^{2}\left(
p+1/2\right) \Gamma \left( 2k-p+1/2\right) }{\left( 2k+1\right) !\sqrt{%
\left( 2p+1\right) !}}\times  \notag \\
&&\left( 2p+1\right) \left[ 2+4k-2p\right] \text{ .}  \label{Tkkp}
\end{eqnarray}%
According to Eqs.~(\ref{w2})-(\ref{Tkkp}) we finally obtain

\begin{equation}
\omega _{k}^{(2)(1/2D)}=\Lambda _{1D}^{2}\left( \frac{\gamma _{k}^{\prime }}{%
\pi ^{2}}-C_{1/2}\right) \text{ ,}
\end{equation}%
with $C_{1/2}=-\frac{3}{4\pi }\left[ -\frac{3}{2}+\frac{7\sqrt{3}}{9}+9\ln 2%
\sqrt{\left( 2-\sqrt{3}\right) }\right] $

\begin{multline}
\gamma _{k}^{\prime }=\frac{4}{\sqrt{\pi }}\sum\limits_{p\neq 0}\left[ \frac{%
\left( -1\right) ^{p+1}(\frac{3}{2}-p)\Gamma ^{2}\left( p+1/2\right) }{%
2^{2p+1}\left( 2k+1\right) !}\times \right. \\
\left. \frac{\Gamma \left( 2k-p+1/2\right) \left( 2p+1\right) \left(
1+2k-p\right) }{pp!}\right] \\
-\sum\limits_{p\neq 0;\text{ }m\neq k}\left[ \left( \frac{1}{m+k}+\frac{4}{%
m-k}\right) \times \frac{\Gamma ^{2}(p+\frac{1}{2}+k)}{\left( 2k+1\right) !}%
\right. \\
\left. \frac{1}{\left( 2p+1\right) !}\left[ \frac{3}{4}-\left( k-p\right)
^{2}+k+p\right] ^{2}\right] \\
-\frac{1}{4}\frac{\Gamma ^{2}(2k+\frac{1}{2})\left( \frac{3}{4}+2k\right)
^{2}}{2k\left[ (2k+1)!\right] ^{2}}\text{ .}
\end{multline}%
For an evaluation of the concentration $c_{k}^{(1/2D)}$ we just need to
substitute in Eqs.~(\ref{C1D}) and (\ref{delC1D}) $n_{0}^{(1D)}(z)%
\rightarrow n_{0}^{(1/2D)}(z),$ $\omega _{k}^{(1D)}\rightarrow \omega
_{k}^{(1/2D)},$ $\varphi _{k}\rightarrow \varphi _{k}^{(1/2)}$ and $%
T_{00kp}\rightarrow \overline{T_{00kp}}$.


\begin{thebibliography}{99}
\bibitem{amo} A. Amo, D. Sanvitto, F. P. Laussy, D. Ballarini, E. del Valle,
M. D. Martin, A. Lema\^{\i}tre, J. Bloch, D. N. Krizhanovskii, M. S.
Skolnick, C. Tejedor, and L. Vi\~{n}a, Nature \textbf{457}, 291 (2009).

\bibitem{Keeling} J. Keeling and N. G. Berloff, Nature \textbf{457},
273(2009).

\bibitem{Sarchi} D. Sarchi and V. Savona, Phys. Rev. B \textbf{77}, 045304
(2008).

\bibitem{Sanvitto} D. Sanvitto, F. M. Marchetti, M. H. Szyma\'{n}ska, G.
Tosi, M. Baudisch, F. P. Laussy, D. N. Krizhanovskii, M. S. Skolnick, L.
Marrucci, A. Lema\^{\i}tre, J. Bloch, C. Tejedor and L. Vi\~{n}a, Nature
Phys. \textbf{6}, 527 (2010).

\bibitem{Lagoudakis} K. G. Lagoudakis, T. Ostatnick\'{y}, A. V. Kavokin, Y.
G. Rubo, R. Andr\'{e}, and B. Deveaud-Pl\'{e}dran, Science \textbf{326}, 974
(2009).

\bibitem{Liew} T. C. H. Liew, M. M. Glazov, K.V. Kavokin, I. A. Shelykh, M.
A. Kaliteevski, and A.V. Kavokin, Phys. Rev. Lett. \textbf{110}, 047402
(2013).

\bibitem{tosi} G. Tosi, G. Christmann, N. G. Berloff, P. Tsotsis, T. Gao, Z.
Hatzopoulos, P. G. Savvidis, and J. J. Baumberg, Nature Physics \textbf{8},
190 (2012).

\bibitem{Bogoliubov} N. N. Bogolyubov, J. Phys. USSR \textbf{11}, 23(1947).

\bibitem{Anderson} M. H. Anderson, et al.,Science \textbf{269}, 198 (1995).

\bibitem{Utsunomiya} S. Utsunomiya, L. Tian, G. Roumpos, C. W. Lai, N.
Kumuda, T. Fujisawa, M. Kuwata-Gonokami, A. L\"{o}ffler, S. H\"{o}fling, A.
Forchel, and Y. Yamamoto, Nature Physics, \textbf{4}, 702 (2008).

\bibitem{Bajoni} D. Bajoni, E. Peter, P. Senellart, J. L. Smirr, I. Sagnes,
A. Lema\^{\i}tre, and J. Bloch, Appl. Phys. Lett. \textbf{90}, 051107 (2007).

\bibitem{Wertz} E. Wertz, L. Ferrier, D. D. Solnyshkov, R. Johne, D.
Sanvitto, A. Lema\^{\i}tre, I. Sagnes,R. Grousson, A. V. Kavokin, P.
Senellart, G. Malpuech and J. Bloch, Nature Physics \textbf{6}, 860 (2010).

\bibitem{Amo1} A. Amo, S. Pigeon, C. Adrados, R. Houdr\'{e}, E. Giacobino,
C. Ciuti, and A. Bramati, Phys. Rev. B \textbf{82}, 081301(R) (2010).

\bibitem{tim} M. Wouters, T. C. H. Liew, and V. Savona, Phys. Rev. B \textbf{%
82}, 245315 (2010).

\bibitem{KavovinPolariton} A. Kavokin, Nature Physics \textbf{8}, 183 (2012).

\bibitem{prb} C. Trallero-Giner, T. C. H. Liew, and A. V. Kavokin, Phys.
Rev. B \textbf{82}, 165421 (2010).

\bibitem{Ciuti} C. Ciuti, V. Savona, C. Piermarocchi, A. Quattropani, and P.
Schwendimann, Phys. Rev. B \textbf{58}, 7926 (1998).

\bibitem{pitaevskii} L. Pitaevskii and S. Stringari, \textit{Bose-Einstein
Condensation} (Clarendon Press, Oxford, 2003).

\bibitem{you1997} L. You, W. Hoston, and M. Lewenstein, Phys. Rev. A \textbf{%
55}, R1581 (1997).

\bibitem{stringari1996} S. Stringari, Phys. Rev. Lett. \textbf{77}, 2360
(1996).

\bibitem{prb2} Y. N\'{u}\~{n}ez Fern\'{a}ndez, M. I. Vasilevskiy, C.
Trallero-Giner, and A. Kavokin, Phys. Rev. B \textbf{87}, 195441 (2013).

\bibitem{GRTable} I. S. Gradshteyn and I. M. Ryzhik, \textit{Tables of
Integrals, Series and Products} (Academic, New York, 1980).

\bibitem{Kohn} G. W. Gibbons and C. N. Pope, Ann. Phys. \textbf{326}, 1760
(2011).

\bibitem{edwards1996} M. Edwards, P. A. Ruprecht, K. Burnett, R. J. Dodd,
and C. W. Clark, Phys. Rev. Lett. \textbf{77} 1671 (1996).

\bibitem{spectr1} H. Mathieu, P. Lefebvre, J. Allegre, B. Gil, and A.
Regreny, Phys. Rev. B \textbf{36}, 6581 (1987).

\bibitem{abo} J. R. Abo-Shaeer, C. Raman, J. M. Vogels, and W. Ketterle,
Science \textbf{292}, 476 (2001).

\bibitem{nota0} Tomas-Fermi limit becomes good approach for the GPE if the
dimensionless parameter $\Lambda \geq 25$ (see Refs.~\onlinecite{prb} and %
\onlinecite{prb2}). For the GaAs/AlAs microcavities this value can be
achieved if the exciton-photon detuning parameter $\delta \approx 0$ meV and
$\mathcal{N}$ larger than $2.5\times 10^{5}.$

\bibitem{spectr2} T. P. Pearsall, Appl. Phys. Lett. \textbf{60}, 1712 (1992).

\bibitem{spectr3} E. D. Kim, A. Majumdar, H. Kim, P. Petroff, and J.
Vuckovic, Appl. Phys. Lett. \textbf{97}, 053111 (2010).

\bibitem{Collett} M.~J. Collett and C.~W. Gardiner, Phys. Rev. A \textbf{30}%
, 1386 (1984).

\bibitem{Walls_book} D.~F. Walls and G.~J. Milburn, \textit{Quantum Optics}
(Springer-Verlag, Berlin 1995)

\bibitem{Nota1} See Refs.~\onlinecite{Rolsi} for general discussion.

\bibitem{Rolsi} A. E. Muryshev, G. V. Shlyapnikov, W. Ertmer , K. Sengstock,
and M. Lewenstein, Phys. Rev. Lett. \textbf{89}, 110401 (2002); L.
Khaykovich and B. A. Malomed, Phys. Rev. Lett. \textbf{74}, 023607 (2006);
C. Trallero-Giner, R. Cipolatti, and T. C. H. Liew, Eur. Phys. J. D \textbf{%
67}, 143 (2013).

\bibitem{Nota} Notice that in Refs. \onlinecite{Wertz} and \onlinecite{tim}
for the solution of 1D equations the authors employed a 2D value for
polariton-polariton interaction constant $\lambda $. This is equivalent to
normalized the 1D wave funciton to width of the wire, $L_{y}$, which leads
to inacurate value of the 1D polariton-polariton interaction constant $g_{1}=%
\mathcal{N}\lambda /L_{y}.$

\bibitem{trallero1} C. Trallero-Giner, Victor Lopez-Richard, Ming-Chiang
Chung, and Andreas Buchleitner, Phys. Rev. A \textbf{79}, 195441 (2009).

\bibitem{trallero2} C. Trallero-Giner, V. L\'{o}pez-Richard, Y. N\'{u}\~{n}%
ez-Fern\'{a}ndez, M. Oliva, G.E. Marques, and M.C. Chung, Eur. Phys. J. D
\textbf{66}, 177 (2012).

\bibitem{Ferrier} L. Ferrier, E. Wertz, R. Johne, D. D. Solnyshkov, P.
Senellart, I. Sagnes, A. Lema\^{\i}tre, G. Malpuech, and J. Bloch, Phys.
Rev. Lett. \textbf{106}, 126401 (2011).

\bibitem{Abramowitz} Handbook of Mathematical Functions, edited by \textit{%
M. Abramowitz and I. Stegun }(Dover, New York, 1972).
\end{thebibliography}
\end{document}